\documentclass[
    reprint,
    aps,
    pra,
    superscriptaddress,
    nofootinbib,
    floatfix
]{revtex4-2}

\usepackage{amsmath}
\usepackage{amssymb}
\usepackage{bm}
\usepackage{graphicx}
\usepackage{dcolumn}
\usepackage{physics}
\usepackage{xcolor}
\usepackage[hidelinks]{hyperref}

\begin{document}

\title{Seeded SU(1,1) interferometry for Fourier-domain optical coherence tomography}

\author{Alejandra A. Padilla}
\email{alejandra.padilla@icfo.eu}
\affiliation{ICFO -- Institut de Ciencies Fotoniques,
The Barcelona Institute of Science and Technology,
08860 Castelldefels, Barcelona, Spain}

\author{Valentina Gacha}
\affiliation{ICFO -- Institut de Ciencies Fotoniques,
The Barcelona Institute of Science and Technology,
08860 Castelldefels, Barcelona, Spain}

\author{Daniel F. Urrego}
\affiliation{ICFO -- Institut de Ciencies Fotoniques,
The Barcelona Institute of Science and Technology,
08860 Castelldefels, Barcelona, Spain}

\author{Juan P. Torres}
\affiliation{ICFO -- Institut de Ciencies Fotoniques,
The Barcelona Institute of Science and Technology,
08860 Castelldefels, Barcelona, Spain}
\affiliation{Department of Signal Theory and Communications,
Universitat Politecnica de Catalunya,
08034 Barcelona, Spain}

\date{\today}

\begin{abstract}
We demonstrate Fourier-domain optical coherence tomography (FD-OCT) based on a seeded SU(1,1) interferometer. Multilayer objects are probed with broadband light centered at $1550~\text{nm}$, while depth-resolved 3D images are reconstructed from photon flux measurements centered at $810~\text{nm}$. We show that, in the low parametric-gain regime, seeding increases the photon flux, enabling volumetric imaging with a spectrometer rather than single-photon detectors. Our analysis further shows that, under the conditions considered, seeding provides a more effective route to sensitivity enhancement than increasing the parametric gain.
\end{abstract}

\maketitle
\section{Introduction}
Optical coherence tomography (OCT) is a three-dimensional imaging technique that provides high resolution images of transparent and semi-transparent samples by exploiting optical coherence in an interferometer~\cite{huang1991OCT, dresel1992OCT}. The axial and transverse resolutions of OCT are independent. To obtain information in the transverse direction (plane perpendicular to the optical beam propagation), OCT focuses light to a small spot that is scanned across the sample. To obtain information in the axial direction (along the optical beam propagation), OCT light with a short coherence length that enables optical sectioning of the sample by measuring the interference patterns.

Since its introduction in the early 1990s, OCT has evolved into a mature and widely used technology~\cite{fujimoto2016development, swanson2017ecosystem, hitzenberger2018optical}, with strong impact in biomedical imaging, most notably in ophthalmology for retinal diagnostics~\cite{langlo2025optical}. OCT has also been applied in many other fields, from material inspection to art conservation~\cite{targowski2012optical}. Most modern OCT implementations rely on Fourier-domain optical coherence tomography (FD-OCT), where depth-resolved images are obtained by taking the Fourier transform of the spectrum of the interference signal~\cite{fercher2003optical}. On the one hand, this approach avoids scanning the reference arm mirror, enabling faster data acquisition compared to time-domain optical coherence tomography (TD-OCT)~\cite{urrego2024non}. On the other hand, FD-OCT can provide a sensitivity enhancement compared to TD-OCT~\cite{leitgeb2003performance}.

In parallel to new technological breakthroughs and new commercial applications, there is interest in exploring alternative approaches to OCT that can provide potential advantages in specific scenarios. The nonlinear process of spontaneous parametric down-conversion (SPDC) can be used to generate broadband light (hundreds of nm) with tunable central frequency~\cite{carrasco2006broadband}, allowing OCT with axial resolutions of $\sim 1~\mu\text{m}$. An active area of research is the development of OCT systems that are sensitive to the presence of specific molecules. One route towards this goal is the use of nonlinear processes such as second-harmonic generation~\cite{jiang2004second} or coherent anti-Stokes Raman scattering (CARS)~\cite{bredfeldt2005molecularly}.

In recent years, there have been several experimental demonstrations of OCT based on two types of interferometers: a Mandel-type induced coherence interferometer~\cite{valles2018optical} and a Yurke-type SU(1,1) interferometer~\cite{paterova2018tunable, machado2020optical, vanselow2020frequency}. These two types of interferometers make use of optical parametric amplifiers at the input and output ports of the interferometer~\cite{yurke19862,mandel1991,mandel1992}, instead of beam splitters, as in conventional interferometers. These interferometers are sometimes referred to as nonlinear interferometers~\cite{chekhova2016nonlinear}.

There are two key motivations for exploring OCT based on SU(1,1) interferometers. First, the sample is probed with light at one wavelength, whereas the detected photons have a different wavelength, i.e., the photons that interact with the sample through reflection/transmission are never detected. This principle has motivated the term "imaging with \textit{undetected photons}"~\cite{lemos2014quantum} for referring to these imaging schemes. This enables the probing wavelength to be selected according to the optical properties of the sample, even when efficient detectors are unavailable in that specific spectral region. The detection wavelength can instead be chosen where {\it efficient} and {\it easily accessible} optical detectors exist. Nonlinear interferometers have been used for methane detection~\cite{florez2022enhanced} by exploiting the strong absorption peak at $3.22\,\mu$m, while detection is performed at $848$~nm. They have been used for Terahertz sensing~\cite{kutas}, where the sample is probed at $1.26$~THz and $0.47$~THz, while detection is performed at $661$~nm.

The second motivation is that SU(1,1) interferometers can show quantum-enhanced sensitivity below the shot-noise limit, where the precision of estimation scales as $ \sim 1/\sqrt{n}$, where $n$ is the number of photons probing the sample, potentially reaching Heisenberg-limited precision in parameter estimation~\cite{marino2012effect,giese2017phase,oglialoro2026below,machado2026improving} where the sensitivity scales instead as $ \sim 1/n$. However, the actual realization of this potential advantage in experimental setups should take into account the detrimental effect of losses, restricting the sensitivity enhancement to selected regions of parameter space \cite{oglialoro2026below}.

Optical parametric amplifiers (OPAs), the defining elements of an SU(1,1) interferometer, are characterized by the parametric gain $G$, which depends on the characteristics of the pump beam and the nonlinear susceptibility $\chi^{(2)}$ of the second-order nonlinear crystal. For CW pump powers of tens of milliwatts, the case of most experimental implementations, it has been shown~\cite{torres2011engineering} that the parametric gain is typically very small, i.e. $G\ll1$. This regime is known as the low parametric gain regime. In this case, single-photon detectors are required, and the low photon flux leads to long acquisition times when high precision imaging is required.

In order to increase the photon flux and reduce the acquisition time, two main approaches can be pursued. The first is to operate the OPAs in the high parametric gain regime ($G \geq 1$). However, this requires the use of a pulsed pump laser with high peak power, adding experimental complexity, cost, and technical challenges. OCT in the high parametric gain regime was demonstrated in 2020~\cite{machado2020optical}, providing high photon flux rates and allowing the use of standard spectrometers for detection, instead of single-photon detectors or highly sensitive CCD cameras. The second alternative is to seed the SU(1,1) interferometer with an external laser to enhance photon flux rates and parameter estimation sensitivity, without relying on high parametric gain. Imaging and spectroscopy based on seeded SU(1,1) interferometers have been demonstrated~\cite{cardoso2018,florez2022enhanced}. More recently, it has been shown that, for phase estimation, seeding can provide a metrological advantage in parameter estimation~\cite{kranias2025metrological}. 

Here we demonstrate Fourier-domain Optical Coherence Tomography (FD-OCT) based on a seeded SU(1,1) interferometer in the low parametric gain regime. We experimentally demonstrate the feasibility of this approach through axial and volumetric imaging of multilayer samples, enabled by the increased photon flux provided by seeding. Beyond this practical advantage, we show that, under the conditions considered, seeding can provide a more effective route to sensitivity enhancement than increasing the parametric gain. In Section~\ref{experimental setup}, we describe the experimental implementation. Section~\ref{experimental results} presents the results of the axial and volumetric images, while Section~\ref{sensitivity enhancement} analyzes the sensitivity advantage of seeding. Section~\ref{conclusions} summarizes the main conclusions of our work.

\section{Experimental setup}
\label{experimental setup}

\begin{figure*}[t]
    \centering
    \includegraphics[width=0.9\textwidth]{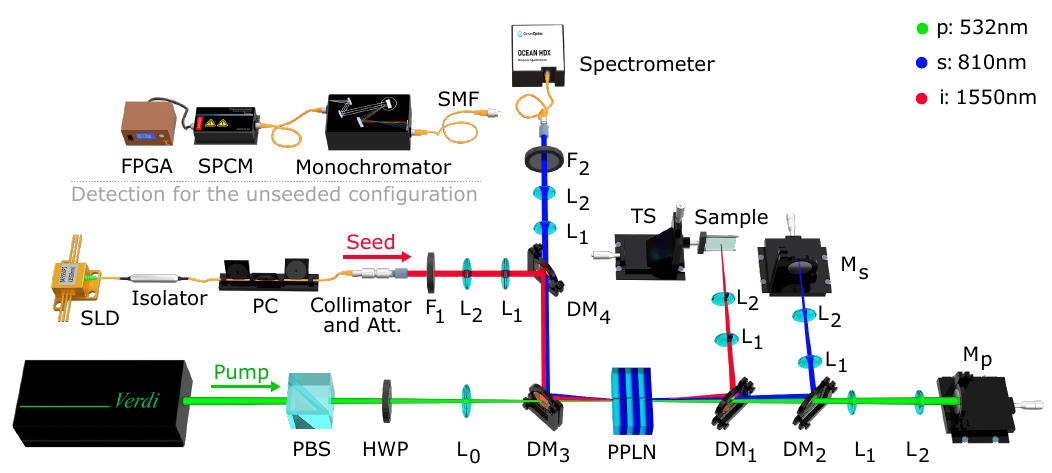}
    \caption{Experimental setup to demonstrate Fourier-domain optical coherence tomography using a seeded SU(1,1) interferometer. Signal ($s_1$) and idler ($i_1$) photons are generated via non-degenerate parametric down-conversion. The idler photons probe the sample. Photons reflected from the sample, together with the reflected pump ($p$) and signal photons, are re-injected into the nonlinear crystal. The interferometer can operate in two configurations: (a) unseeded, where detection is performed using single-photon detectors, and (b) seeded, where a broadband superluminescent diode (SLD) increases the photon flux rate, enabling detection with a spectrometer. PBS: polarizing beam splitter; HWP: half-wave plate; DM: dichroic mirror; $L$: lens; F: frequency filter; M: mirror; TS: translation stage; SPCM: single-photon counting module; PC: polarization controller; and SMF: single mode fiber.}
    \label{fig:setup}
\end{figure*}

The experimental implementation of the seeded SU(1,1) interferometer used for FD-OCT is shown in Fig.~\ref{fig:setup}. The CW pump beam at $\lambda_p = 532~\text{nm}$, with a linewidth of $5~\text{MHz}$, is focused using a lens ($\text{L}_{\text{0}}$) to a beam waist of $\text{w}_p = 75~\mu\text{m}$. The beam waist is located at the center of a periodically poled lithium niobate (PPLN) crystal with length $L=1~\text{mm}$. The polarization of the pump beam is controlled using a polarizing beam splitter (PBS) and a half-wave plate (HWP). Signal $s_1$ and idler $i_1$ photons are generated via non-degenerate type-0 parametric down-conversion, with central wavelengths of $\lambda_s=810~\text{nm}$ and $\lambda_i=1550~\text{nm}$, respectively.

Since the operation of the interferometer depends on the parametric gain of the nonlinear interaction, we estimate the gain under our experimental conditions. For a narrowband CW pump beam, the parametric gain of the nonlinear process can be estimated as~\cite{torres2011engineering,dayan2007}
\begin{equation}
G_{CW}=\left[ \frac{\omega_s \omega_i \left[ \chi^{(2)} \right]^2 P_{m}}{8 \epsilon_0 c^3 n_p n_s n_i A_{eff}} \right]^{1/2} L
\label{gain}
\end{equation}
where $P_m$ is the mean power of the CW pump, $A_{eff}=2\pi\text{w}_p^2$ is the effective area, $\chi^{(2)}$ is the second-order susceptibility of the nonlinear crystal, $\omega_{p,s,i}$ are the central angular frequencies of all waves involved, and $n_{p,s,i}$ are the corresponding refractive indices obtained from the Sellmeier equations in Ref.~\cite{covesion}. Using the experimental parameters of our setup [pump beam power: $3.5$ mW, nonlinear crystal temperature: $125^{\circ}\text{C}$, and $\chi^{(2)}=14$ pm/V], Eq.(\ref{gain}) yields $G_{CW} \sim 7 \times 10^{-5}$. The system therefore operates in the low parametric gain regime.

After the first pass through the nonlinear crystal, the pump, signal, and idler photons are separated using dichroic mirrors $\text{DM}_1$ and $\text{DM}_2$. The pump and signal photons are reflected back into the nonlinear crystal by mirrors $\text{M}_{p,s}$ mounted on translation stages (TS), allowing the control of their phases before reentering the crystal. Meanwhile, the idler photons illuminate the sample, which is also mounted on a translation stage for axial and transverse scanning. The lenses $\text{L}_1 = 400~\text{mm}$ and $\text{L}_2 = 100~\text{mm}$ ensure mode matching of the returning photons in each of the arms of the interferometer and provide a focused spot size of $\approx 18.5~\mu\text{m}$ in the sample plane.

The reflected pump, signal, and idler photons are recombined in the nonlinear crystal, where the process of parametric amplification generates new signal-idler pairs ($s_2$, $i_2$). The output signal photons $s_2$ are separated from the pump by $\text{DM}_3$ and $\text{DM}_4$, coupled into a single-mode fiber, and spectrally filtered by filter $\text{F}_2$ to suppress residual pump power and background light.

The setup can be switched between seeded and unseeded configurations by turning on or off a superluminescent diode (SLD) centered at $1550~\text{nm}$, with a spectral width of $40~\text{nm}$, determined by the bandwidth of frequency filter $\text{F}1$. Supplement 1 compares the spectra of the SPDC photons in the unseeded configuration and the spectrum of the SLD. Although their central wavelengths differ slightly, they exhibit a strong spectral overlap, resulting in a smooth Gaussian-like spectrum. This can be explained by the Gaussian product theorem~\cite{wolf1995}, that states that the overlap of two Gaussian functions with different widths and centers is still a Gaussian function.

In the unseeded configuration, detection of the extremely low photon flux-rate of signal photons requires the use of a monochromator and a single-photon counting module (SPCM), with data acquisition performed using a field-programmable gate array (FPGA). In the seeded configuration, light from the SLD is injected into the interferometer. Back-reflections are suppressed using a fiber isolator, while a polarization controller (PC) and variable attenuators (Att.) adjust the polarization and power. Lenses $\text{L}_1$ and $\text{L}_2$ in the seed path provide the required mode matching between seeding light and the pump beam at the nonlinear crystal. The increased signal photon flux provided by seeding enables detection with a spectrometer and real-time acquisition of spectral interferograms for Fourier-domain reconstruction of the depth-resolved OCT signal.

In FD-OCT based on an SU(1,1) interferometer, the spectrum of the output signal photons $s_2$ is measured. In Supplement 2 we derive in detail the theoretical expression of the spectrum of the output signal photons as a function of the experimental parameters and the sought-after frequency-dependent reflectivity $r_i(\Omega)$ of the sample. Accurate reconstruction of the depth profile from this spectrum requires the optical path-length difference between the interferometer arms to lie within a specific range, determined by the source bandwidth and the spectral resolution of the detector. It is shown in~\cite{machado2020optical}, and derived in detail in Supplement 3, that the correct operation of FD-OCT requires \begin{equation}
\frac{\lambda_s^2}{\Delta\lambda_s}
\ll
\Delta L+N_g L
\ll
\frac{\lambda_s^2}{\delta\lambda_s}
\end{equation}
Here $\lambda_{s}$ is the central wavelength of signal photons, $\Delta \lambda_s$ is their bandwidth, $\delta \lambda_s$ is the spectrometer resolution, $L$ is the nonlinear crystal length, $\Delta L$ is the optical path-length difference between the signal and idler arms of the interferometer, and $N_g=n_{g,s}-n_{g,i}$ is the difference between the group indices of the signal and idler photons in the nonlinear crystal. Since the pump beam is a narrowband CW source, we have $\lambda_s^2/\Delta\lambda_s=\lambda_i^2/\Delta\lambda_i$. In our experiments, $\lambda_s=810~\text{nm}$, $\Delta \lambda_s = 17.4~\text{nm}$, and the spectrometer resolution is $\delta \lambda_s=0.36~\text{nm}$. Using the Sellmeier equations, $n_{s,i}$, we obtain that $N_g L=77.6\,\mu$m. Therefore, the optical path-length difference, $\Delta L$, controlled by a translation stage, must satisfy $ 40~\mu\mathrm{m} \ll \Delta L + N_g L \ll 1.72~\mathrm{mm}$. The left inequality ensures that the spectral fringes are sufficiently dense to encode depth information. In our implementation, the group-delay offset $N_gL$ already exceeds the lower bound, so this condition is automatically fulfilled. Consequently, the experimental alignment consists of adjusting $\Delta L$ to keep the total delay below the upper bound, ensuring that the fringe spacing remains resolvable by the spectrometer.

\section{Experimental results}
\label{experimental results}

\begin{figure*}[]
    \centering
\includegraphics[width=0.9\linewidth]{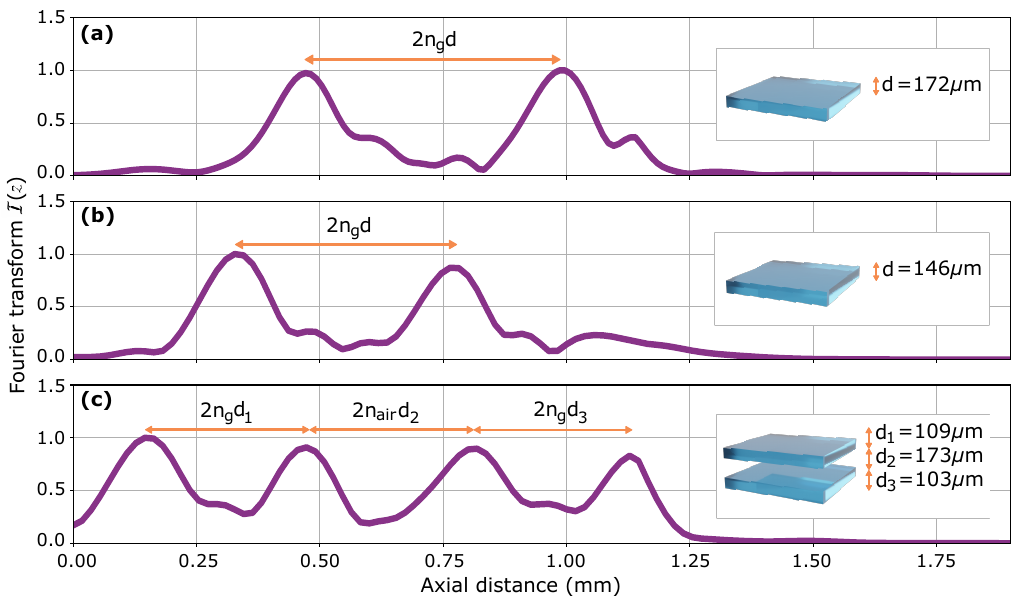}
    \caption{Fourier transform ${\cal I}(z)$ of the spectrum signal photons for single-layer and multilayer glass samples using a seeded SU(1,1) interferometer. Peaks correspond to reflections from refractive-index discontinuities within the samples. The reported thicknesses $d$ are extracted from the peak separations using $\Delta z = 2 n_g d$, with $\text{n}_g=1.5$. Manufacturer-specified thicknesses are: (a) $170~\mu\text{m}$, (b) $130-160~\mu\text{m}$, and (c) multilayer structure of $85-115~\mu\text{m}$ glass, $130-160~\mu\text{m}$ air gap, and $85-115~\mu\text{m}$ glass. Insets show schematic representations of the samples.}
    \label{fig:sample_glass}
\end{figure*}

\subsection{Axial scan of multilayer samples using seeding}

To validate the performance of the seeded SU(1,1) interferometer for OCT, we first measured uncoated glass samples with different thicknesses. At the idler wavelength ($1550~\text{nm}$), uncoated glass reflects approximately $4\%$ of the incident light per interface, so the visibility of the interference pattern is highly reduced. In the low parametric gain regime, with no seeding, the low photon flux rates typically require the use of single-photon sensitive detection schemes. Seeding the interferometer increases the detected photon flux, enabling spectral acquisition with a spectrometer instead of single-photon detection. For the measurements described here, the pump power is $3.5~\text{mW}$, the seed power is $20~\text{mW}$ and the acquisition time is 100$\mu\text{s}$.

Figure~\ref{fig:sample_glass} shows the reconstructed axial scans (A-scans). At each measurement point, two spectra, $S(\lambda_s)$, were recorded with a relative phase shift of $\pi$. Their subtraction suppresses the DC contribution and the autocorrelation terms, leaving only the interference signal associated with the reflections from the sample. The resulting spectrum is rewritten as a function of the wavenumber $k = 2\pi/\lambda$. Considering the Jacobian of the transformation, we obtain $S(k)=(2 \pi/k^2)\, S\left(\lambda\right)$. The spectrum is then resampled to obtain a function $S(k)$ with equally spaced $k$-values. Finally, we obtain the Fourier transform of ${\cal I}(z)={\cal F} \left[ S(k) \right]=\int dk\, S(k)\,\exp \left( i k z \right)$. 

In FD-OCT, the detected signal amplitude decreases with increasing optical path difference due to the finite spectral resolution of the detection system. This effect, known as {\it sensitivity decay}, leads to a systematic attenuation of deeper features in the reconstructed depth profile~\cite{leitgeb2003performance}. In Supplement 4 we describe the method to correct for the {\it sensitivity decay}. We will apply the correction to all OCT scans presented here.

The function ${\cal I}(z)$, plotted as a function of the axial distance $z$, is shown in Fig.~\ref{fig:sample_glass}. Each peak corresponds to an optical interface arising from refractive-index discontinuities within the sample. The distance $\Delta z$ between the positions of each peak encodes the optical path length difference $\Delta z= 2 n_g d$, where $d$ is the thickness of the layer and $n_g$ is the group index of the material that makes the layer. Since $n_g~\approx 1.5$, the shape of ${\cal I}(z)$ enables reconstruction of the internal layer composition of the sample.

Fig.~\ref{fig:sample_glass}(a) and (b) correspond to two single-layer glass samples with different thickness, with nominal values of $172\, \mu$m and $146\, \mu$m. ${\cal I}(z)$ exhibits two dominant peaks corresponding to the front and back surfaces of the single-layer sample. In Fig.~\ref{fig:sample_glass}(c) we consider two glass samples separated by a distance of $173\,\mu$m, which corresponds to a multilayer sample with three layers. We observe four peaks that correspond to reflections from the front and back surfaces of each glass layer, separated by an air gap. Our results are in good agreement with manufacturer specifications. These experimental results validate the ability of the seeded SU(1,1) interferometer to resolve axial structures in weakly reflecting transparent media.

\begin{figure*}[t]
    \centering
\includegraphics[width=1\textwidth]{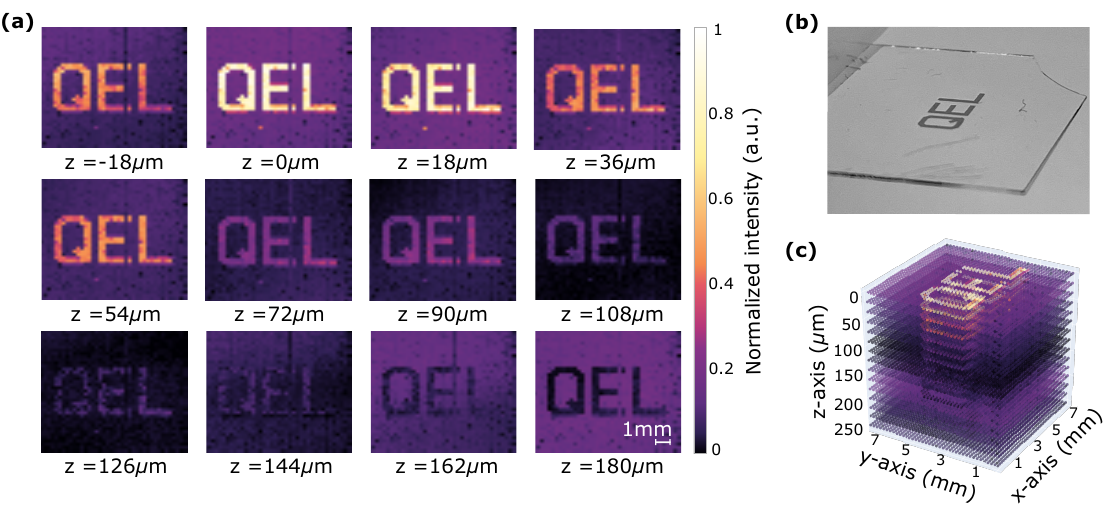}
    \caption{Volumetric reconstruction of a $170~\mu\text{m}$ thick glass sample with a $300~\text{nm}$ aluminum pattern deposited on the top surface, measured with a seeded SU(1,1) interferometer. (a) Cross-sectional view of the reconstructed volume, showing the strong reflections from the aluminum and weaker reflections from the glass surface. (b) Photograph of the sample. (c) Three-dimensional reconstruction obtained by stacking the A-scans ${\cal I}(z)$.}
    \label{fig:QEL}
\end{figure*}

\subsection{Volumetric OCT imaging with a seeded SU(1,1) interferometer} 

We now demonstrate volumetric FD-OCT imaging using the seeded SU(1,1) interferometer. The sample consists of a $170~\mu\text{m}$ thick glass layer with a $300~\text{nm}$ aluminum pattern deposited on the top surface (see Supplement 5 for fabrication details). The pump and seed powers are $3.5~\text{mW}$ and $20~\text{mW}$, respectively. The beam is focused onto the sample and raster-scanned with $0.2~\text{mm}$ steps in both transverse directions. At each position, the signal spectrum is recorded and Fourier transformed to obtain an A-scan ${\cal I}(z)$. Stacking all A-scans produces a three-dimensional reconstruction of the sample.

Figure~\ref{fig:QEL}(a) shows a cross-sectional view of the reconstructed volume. The plane z=0 is chosen to coincide with the maximum of the first peak of the A-scan, corresponding to the front surface of the sample, where the aluminum pattern produces a strong reflection revealing the letters QEL. At $1550~\text{nm}$, aluminum reflects approximately $97\%$ of the light, compared to $4\%$ of uncoated glass, resulting in high-contrast reflection between coated and uncoated regions. A second peak appears at $z\approx 180\mu m$, corresponding to the back surface of the glass sample. Since no aluminum is present at this interface, the letters appear as regions of reduced intensity.

Figure~\ref{fig:QEL}(c) shows the three-dimensional reconstruction obtained by stacking the reconstructed A-scans. A custom color map was designed to highlight the weak reflections from the glass and the strong reflections given by the aluminum pattern. The gradual fading of the QEL pattern along the $z$ axis reflects the Gaussian-shaped axial response of ${\cal I}(z)$. The displayed axial range extends beyond the physical sample to show the full reconstructed response. A photograph of the sample is shown in Fig.~\ref{fig:QEL}(b) for reference.These results demonstrate that the seeded SU(1,1) interferometer enables true volumetric OCT imaging while preserving sufficient contrast to distinguish highly reflective metallic features from weakly reflective transparent interfaces.

\subsection{The effect of seeding on photon flux rates and image acquisition time}

\begin{figure*}[]
    \centering
    \includegraphics[width=1\linewidth]{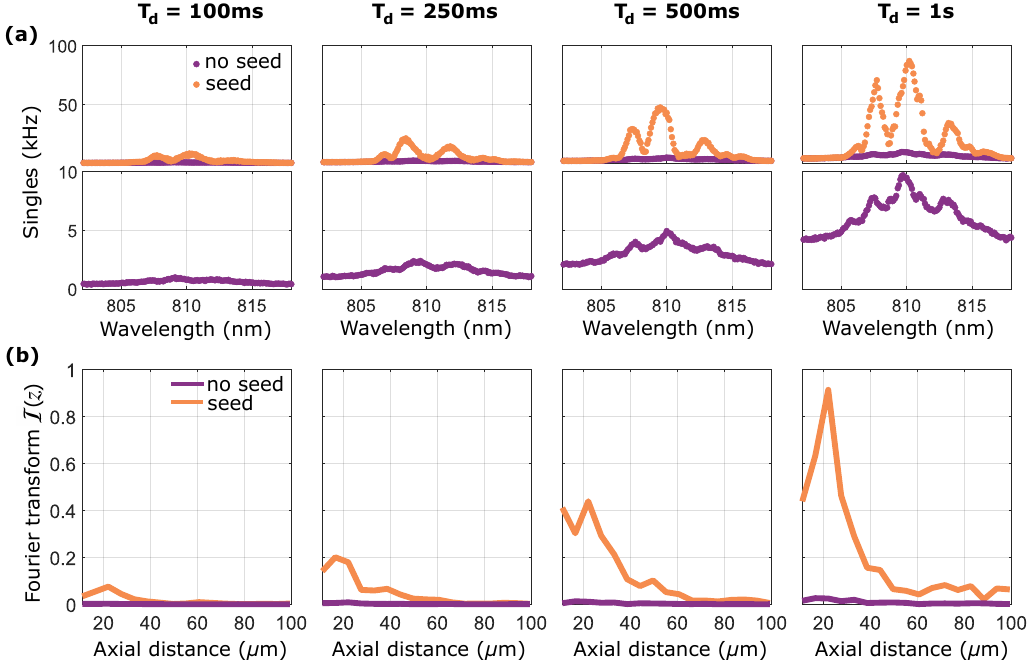}
    \caption{Comparison of signal spectra and A-scans when a silver mirror is used as sample. We consider a seeded (orange) and unseeded (purple) SU(1,1) interferometer for different acquisition times $T_d$. (a) Raw signal spectra at $810~\text{nm}$. The top row shows a comparison between the two configurations, and the bottom row shows a magnified view of the unseeded signal. (b) Corresponding A-scans (Fourier transforms) of the spectra shown in (a). The position of the single peak corresponds to the optical path difference between the signal/idler interferometer arms, mapping the mirror position in FD-OCT.}
    \label{fig:seedVSnoseed}
\end{figure*}

To quantify the effect of seeding on the photon flux rates and image acquisition time, we perform FD-OCT using a silver mirror as a single-layer sample. We compare the signal spectrum obtained with and without seeding under otherwise identical experimental conditions. The signal spectrum at $810~\text{nm}$ was recorded using a monochromator and detected with a SPCM. The pump power was fixed to $9.2~\text{mW}$ for all measurements. For the seeded configuration, the seed power is set at $12~\mu\text{W}$.

Measurements are performed for different acquisition times, $T_d$= 100~\text{ms}, 250~\text{ms}, 500~\text{ms}, 1\text{s}. For each acquisition time, the spectra are recorded sequentially with the seed on and off, ensuring identical experimental conditions. Fig.~\ref{fig:seedVSnoseed}(a) shows the measured signal spectra for both configurations. The top row compares the seeded and unseeded measurements, while the bottom row shows a magnified view of the unseeded spectra. 

A clear increase in the detected photon flux rate is observed when the interferometer is seeded. For all acquisition times considered, the seeded configuration yields an increase of approximately one order of magnitude compared to the unseeded case. In the absence of seeding, increasing $T_d$ leads to a proportional increase of both the signal, while the seeded configuration provides a significantly higher signal level even for short acquisition times. 

The corresponding A-scans obtained from the Fourier transform of the spectra are shown in Fig.~\ref{fig:seedVSnoseed}(b). As expected, for a mirror sample, a single peak is observed, and its position corresponds to two times the optical path difference between the signal/idler interferometer arms. In the seeded configuration, the mirror position can be already retrieved for the shortest acquisition time, $T_d=100~\text{ms}$. In contrast, under the same experimental conditions, the unseeded configuration requires longer acquisition times to achieve a comparable visibility. These results demonstrate that even a low power seed substantially increases the detected photon flux rates, enabling faster detection.

\section{Comparison of routes for sensitivity enhancement in SU(1,1) configurations}
\label{sensitivity enhancement}

In Supplement 6, we show in detail how to calculate the precision with which the reflectivity $R_i$ of s single layer sample can be estimated when we measure the flux rate of signal photons $s_2$. For the sake of simplicity, we make use of the single-mode approximation. For a CW pump, the number of signal photons $s_2$ detected during an acquisition time $T_d$ can be written as $S_{CW}= \langle N_{s_2} \rangle\,\Delta f\, T_d$, where $\langle N_{s_2} \rangle$ is the spectra flux density (photons/s/Hz) obtained in the single-mode approximation and $\Delta f$ is the effective detection frequency bandwidth~\cite{gene2006}. 

\begin{figure*}[t!]
    \centering
    \includegraphics[width=1\linewidth]{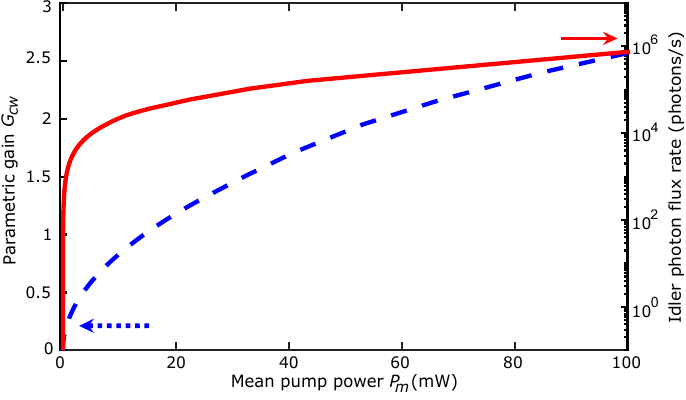}
    \caption{Parametric gain (left axis) and generated photon flux rate of idler photons (right axis) as function of mean pump power.}
    \label{fig:gain_vs_power}
\end{figure*}

The precision of parameter estimation can be obtained as $\sigma_{CW}=\sigma/\sqrt{\Delta f\, T_d}$, where 
\begin{equation}
\sigma= \frac{\sqrt{\text{Var} \left( N_{s_2} \right)}}{\Big| \partial \langle N_{s_2} \rangle/\partial R_i \Big| }
\end{equation}
is the precision obtained in the single mode approximation (see Supplement 6). For the case of a pulsed pump beam, one can obtain $S_{p}= \langle N_{s_2} \rangle\,D\,\Delta f\, T_d$ and $\sigma_{p}=\sigma/\sqrt{D\, \Delta f\, T_d}$ where $D$ is the duty cycle of the pulsed laser~\cite{gene2006}.

The simplest configuration for OCT based on a nonlinear interferometer uses a CW pump without seeding of signal/idler photons. According to Eq.(\ref{gain}), for pump powers of up to tens of mW, this implies that this configuration works in the low parametric gain regime. OCT in this regime has been demonstrated using a Mandel-type induced coherence configuration~\cite{valles2018optical}, as well as a Yurke-type SU(1,1) interferometer in both the time domain~\cite{paterova2018tunable} and Fourier-domain~\cite{vanselow2020frequency}.

To improve the sensitivity and reduce the acquisition time, the flux rate of idler photons that probe the sample must be increased. One option is to increase the pump beam power to values of many hundreds of milliwatts while remaining in the low parametric gain regime. For example, one can achieve signal/idler flux rates of up to $1.5\,\mu$W in a bandwidth of $\sim 30$ nm~\cite{DayanSilberberg2005}. However, this approach requires a high power pump laser and might not be suitable for nonlinear crystals with a lower damage threshold. 

Another option is to work in the high parametric gain regime. However, reaching this regime with the same mean pump power requires a pulsed laser instead of a CW pump. When using a pulsed pump laser, the parametric gain $G_p$ is related to the CW gain $G_{CW} $ by $G_p=G_{CW}/\sqrt{D}$, where $G_{CW}$ is given by Eq.(\ref{gain}) and $D=f_r T_p$ is the duty cycle of the pulsed laser. In~\cite{machado2020optical}, they demonstrated Fourier-domain OCT in an SU(1,1) interferometer in the high parametric gain regime, with a gain of $G_p=1.7$, generating $1.6$ pW of idler photons. The PPLN crystal was pumped with a Nd:YAG laser generating pulses of duration $T_p=18$ ps and beam waist of $40\,\mu$m, with a repetition rate of $f_r=1$ kHz. For these experimental values, and for the same values of $P_m$, there is an enhancement of the parametric gain in the pulsed case over the CW case of $1/\sqrt{D} \sim 7400$. 

The parametric gain can also be increased by using a longer nonlinear crystal. However, high resolution OCT requires the use of large bandwidths, and it is well known that the bandwidth is reduced when increasing the nonlinear crystal length. Fig.~\ref{fig:gain_vs_power} shows the parametric gain and the flux rate of idler photons probing the sample achievable for pump mean powers of up to $100~\text{mW}$. Notice that under these conditions, one can achieve parametric gains of up to $G_{CW}=2.5$, and generate idler photon flux rates of up to $10^{6}$ photons/s.

\begin{figure*}
    \centering
    \includegraphics[width=1\linewidth]{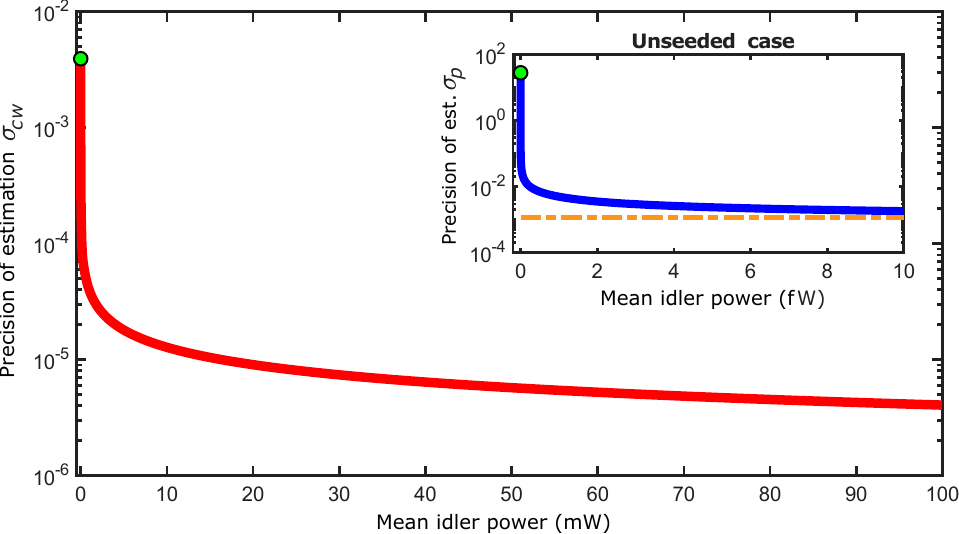}
    \caption{Comparison of the precision in the estimation of the reflectivity $R_i$ for two SU(1,1) configurations. The red curve shows the precision of estimation $\sigma_{cw}$ corresponding to a seeded configuration in the low parametric gain, where we change the CW seed power and keep constant the CW pump power. The parametric gain is $G=10^{-4}$. The blue curve shows the precision of estimation $\sigma_p$ corresponding to an unseeded configuration, where we change the mean pump power of the pulsed laser source. In both cases, we plot the sensitivity as function of the flux rate of idler photons probing the sample. Main panel: estimation precision, $\sigma$, as a function of the idler optical power probing the sample. The dashed line in the inset shows the threshold minimum precision than can be reached in the high parametric gain regime. Parameters: $R_i=0.1$, $R_s=0.8$, and $T_s=0.8$. We consider a detection bandwidth of $\Delta \lambda_s=10$ nm.}
    \label{fig:seed_gain}
\end{figure*}

These considerations motivate an alternative route. OCT in a seeded SU(1,1) interferometer working in the low parametric gain regime allows to increase the flux rate of idler photons probing the sample, and therefore the sensitivity of parameter estimation, while at the same time using {\it easily accessible} CW lasers for pumping and seeding with up to a few mW of power. A question that naturally arise is if the use of seeded SU(1,1) interferometer offers, apart from a practical advantage for its {\it easiness} of experimental implementation, a fundamental advantage, even in principle, when compared with going to the high parametric regime.

Figure~\ref{fig:seed_gain} compares the best precision, i.e., the minimum value of $\sigma_{cw}$ and $\sigma_{p}$, that could be achieved in the estimation of the reflectivity $R_i$ using one of two strategies: i) increasing the mean power of the seed in an SU(1,1) interferometer working in the low parametric regime (red curve) and ii) increasing the mean pump power of a pulsed laser source in an unseeded SU(1,1) interferometer for reaching the high parametric gain regime (blue curve in the inset). The red curve considers a constant parametric gain of $G_{cw}=10^{-4}$ and it starts at $|\alpha|^2=0$ (green dot), corresponding to the seeded configuration, while the blue curve corresponds to the unseeded configuration, and begins at $G_{p}=10^{-4}$ (green dot in the inset).

The comparison shows that seeding, even at the most fundamental level, provides better precision in the estimation of the reflectivity $R_i$. Increasing the seed generates higher flux rates of the idler photons probing the sample and always provides a sensitivity advantage. Using Eq.~(29) of Supplementary 6, we see that the sensitivity scales as $\sigma_{CW} \sim 1/\sqrt{|\alpha|^2 \Delta f T_d}$, where $|\alpha|^2 \Delta f$ is the flux rate of seed photons probing the sample with an effective bandwidth $\Delta f$. In contrast, increasing the parametric gain alone (no seeding) leads to saturation of the achievable sensitivity (orange dashed curve in the inset of Fig.~\ref{fig:seed_gain}). The sensitivity for large gains saturates to the constant value $\sigma_{p}=\sigma_{min}/\sqrt{D \Delta f T_d}$, where $\sigma_{min}$ is given by Eq.~(S30) in Supplementary 6 and $D$ is the duty cycle of the pulsed pump laser. Therefore, while both strategies increase the photon flux probing the sample, only seeding provides a continuously improving sensitivity under the conditions considered.

\section{Conclusions}
\label{conclusions}

We have demonstrated experimentally a Fourier-domain OCT imaging system based on a seeded Yurke-type SU(1,1) interferometer. The main advantage of this OCT configuration is that it allows to use optimum frequency ranges for probing the sample and for the detection stage. This opens many possibilities, especially when there are no efficient or easy-to-use detectors at the frequency range used for probing the sample. This is a severe limiting factor in standard OCT configurations. 

Our experiments join the group of experiments that demonstrate OCT based on nonlinear interferometers with different configurations: time-domain OCT in a Mandel-type induced coherence interferometer~\cite{valles2018optical}, time-domain~\cite{paterova2018tunable} and Fourier-domain~\cite{vanselow2020frequency} OCT in a Yurke-type SU(1,1) interferometer, and Fourier-domain OCT in a Yurke-type SU(1,1) interferometer in the high parametric gain regime~\cite{machado2020optical}. 

One question that naturally arise is what is it the {\it optimum} configuration to do high-resolution OCT, in both the transverse and longitudinal directions, with short detection times. For this, one needs to do Fourier-domain OCT~\cite{leitgeb2003performance,urrego2024non} and to enhance the flux rate of idler photons probing the sample under consideration. The flux rate of idler photons can be increased going to the high-parametric regime. From a fundamental perspective, this regime can potentially allow going beyond the Shot-noise limit of sensitivity, although losses, most of the times unknown with precision in {\it real} experimental setups, should be considered to determine under what conditions this potential advantage can be actually realized in experiments~\cite{oglialoro2026below}.

From a {\it practical} point of view, one might be more interested in enhancing the sensitivity of OCT, and doing it using an {\it easily accessible} experimental setup. We have demonstrated that the use of a seeded SU(1,1) interferometer fulfills both conditions. On the one hand, a seeded interferometer show better precision in the estimation of the value of the reflectivity of an object. Indeed, we have shown that increasing the value of the parametric gain in an unseeded configuration does not provide any sensitivity enhancement beyond a certain value. This is not the case with a seeded configuration, where increasing the power of the seed laser always provides a sensitivity advantage.

On the other hand, we have shown that a seeded configuration can provides sensitivity enhancements using easily available CW lasers delivering average powers of up to a few tens of mW. On the contrary, reaching similar values of the average flux rate in the high parametric regime, it requires the use of a cumbersome and expensive pulsed laser source.

We have shown that OCT with an unseeded configuration generates low photon flux rates, which needs to be detected using single photon counting modules. As a consequence, long detection times are needed to obtain high resolution and high quality images. In contrast, seeding with a broadband laser increases the stimulated photon flux rate, while keeping the axial resolution of the OCT system. The higher photonic flux rates enables spectrometer-based detection and substantially reduces the acquisition time required to retrieve depth-resolved information from weakly reflective samples.

\section{Acknowledgments}
We acknowledge support from the project NOVISLIGHT (PID2023-149780NB-I00) funded by Ministerio de Ciencia, Innovaci\'on y Universidades (Proyectos de generaci\'on de conocimiento 2023). 
This work is part of the R$\&$D project CEX2024-001490-S, funded by MCIN/ AEI/10.13039/501100011033/. We acknowledge support from Fundació Cellex, Fundació Mir-Puig, and from Generalitat de Catalunya through the CERCA program. 

This document has not been peer reviewed. 

\section{Supplementary material}

\setcounter{figure}{0}
\renewcommand{\thefigure}{S\arabic{figure}}

\setcounter{equation}{0}
\renewcommand{\theequation}{S\arabic{equation}}

\section*{Supplement 1: Spectrum of parametric down-conversion and the superluminescent diode}

To characterize the spectral properties of the parametric down-conversion (PDC) process, and to compare it with the spectrum of the injected seed, we measured the signal spectrum around $810~\text{nm}$ using a monochromator. In the unseeded configuration, the signal photons were spectrally resolved by scanning a monochromator and recording the photon flux as a function of wavelength, providing direct access to the PDC spectrum generated in the nonlinear crystal.

\begin{figure*}[t!]
    \centering
    \includegraphics[width=0.9\linewidth]{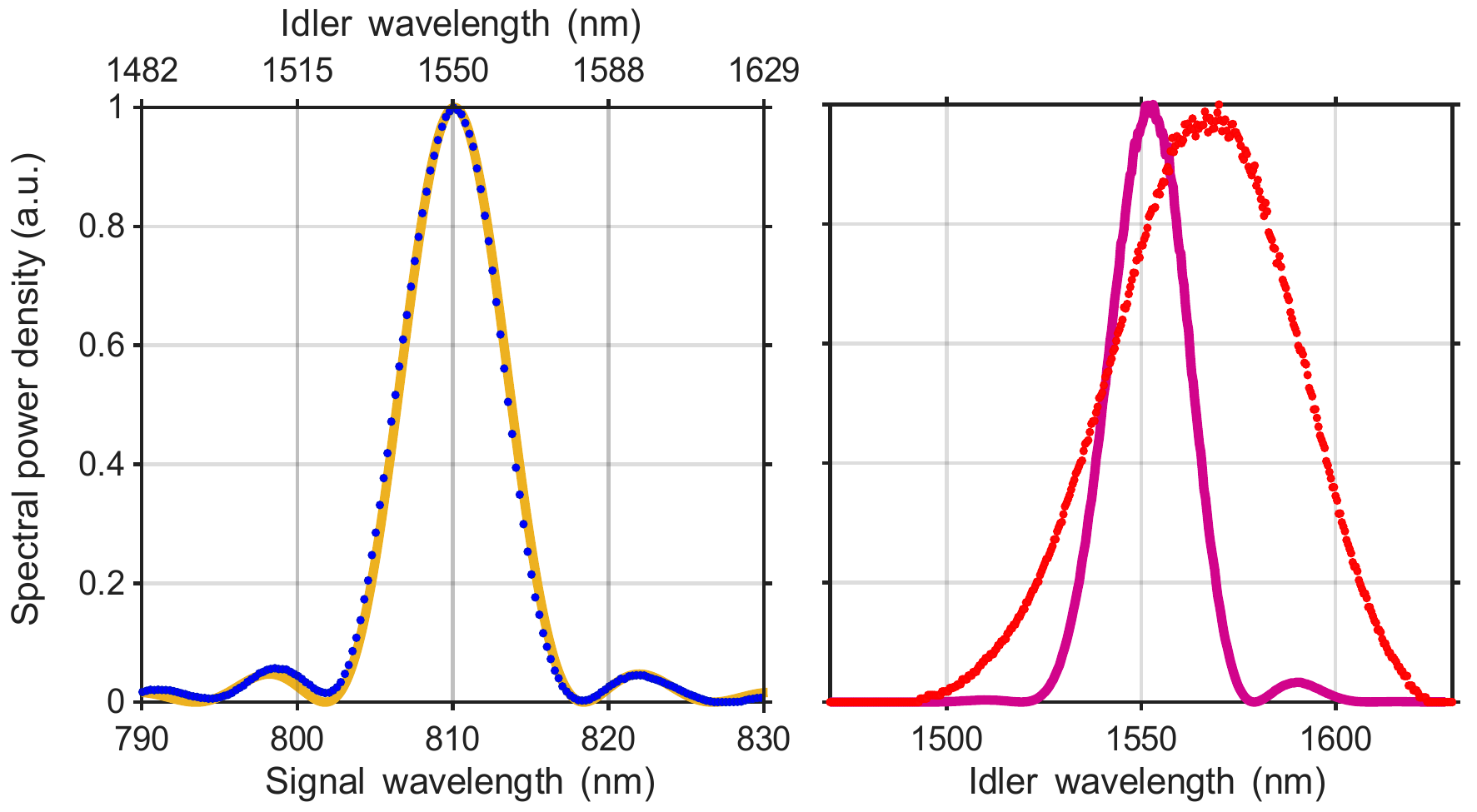}
    \caption{Comparison between the measured SPDC spectrum and the injected seed spectrum.
Left: SPDC signal spectrum centered at 810~nm measured with a monochromator (blue dots), compared with Eq.~\eqref{eq.UV} evaluated for our experimental parameters (solid line). The upper horizontal axis shows the corresponding idler wavelength obtained from energy conservation equation.
Right: Spectrum of the superluminescent diode (SLD) measured with an optical spectrum analyzer (purple dots). The yellow shaded curve corresponds to the product of the measured SLD spectrum and the reconstructed SPDC idler spectrum, illustrating the spectral overlap between the seed and the SPDC bandwidth.}
    \label{fig:SPDC_seed}
\end{figure*}

The corresponding idler spectrum at $1550~\text{nm}$ was obtained indirectly by applying energy conservation,
\begin{equation}
\omega_p = \omega_s + \omega_i 
\end{equation}
using the measured signal spectrum and the known pump frequency. This approach allows reconstruction of the idler spectrum without direct photon-counting detection at telecommunication wavelengths, where losses and detector inefficiencies are more pronounced.

The spectrum of the broadband superluminescent diode (SLD) used for seeding was measured independently using an optical spectrum analyzer (OSA). This measurement enables a direct comparison between the reconstructed idler SPDC spectrum and the spectral bandwidth of the injected seed.

Figure~\ref{fig:SPDC_seed} summarizes these measurements. The left panel shows the measured PDC signal spectrum at $810~\text{nm}$ together with the theoretical prediction derived from the phase-matching conditions of the PPLN crystal. The right panel shows the corresponding idler spectrum translated to $1550~\text{nm}$ via energy conservation, overlaid with the independently measured SLD spectrum. The strong spectral overlap confirms that the seed efficiently addresses the PDC bandwidth relevant for the operation of the SU(1,1) interferometer.

\section*{Supplement 2: Derivation of the spectrum of the output signal photons measured with the spectrometer}
The SU(1,1) interferometer that we consider here makes use of two parametric down-conversion processes (parametric amplification), pumped by a CW pump with frequency $\omega_p$. The two parametric down-conversion processes take place in the same second-order nonlinear crystal, in the forward and backward directions. In the first pass through the nonlinear crystal, the interaction of the pump beam with the molecules of the nonlinear crystal mediates the generation of pairs of signal ($s_1$) and idler $i_1$ photons, with frequencies $\omega_s+\Omega$ and $\omega_i-\Omega$. $\omega_{s,i}$ are the central frequencies and $\Omega$ is the frequency deviation from the corresponding central frequency.

The efficiency of the process is so low that we can assume that the pump beam is un-depleted and can be treated as a classical beam. In the low parametric gain regime of parametric amplification,  that is the case in our experiments, the relationship between input ($b_s$ and $b_i$) and output ($a_{s_1}$ and $a_{i_1}$) quantum operators in parametric down-conversion is described by the Bogoliubov transformations
\begin{eqnarray}
& & a_{s_1}=U_{s_1} b_s(\Omega)+V_{s_1} b_i^{\dagger}(-\Omega) \nonumber \\
& & a_{i_1}=U_{i_1} b_i(\Omega)+V_{i_1} b_s^{\dagger}(-\Omega) \label{bogoliuvov1}
\end{eqnarray}
where for $j\equiv s,i$
\begin{eqnarray}
& & U_{j_1}=\exp \left[ i k_j(\Omega) L\right]  \\
& & V_{j_1}= -i G\, \text{sinc} \left( \frac{\Delta k(\Omega) L}{2} \right) \exp \left[ i k_j(\Omega) L+i\varphi_{p_1}\right] \nonumber
\label{eq.UV}
\end{eqnarray}
$\varphi_{p_1}$ is the phase of the pump beam in the first parametric interaction, $\Delta k (\Omega)=k_p-k_s(\Omega)-k_i(-\Omega)$ is the phase matching function that determines the bandwidth of the parametric interaction, $k_p=\omega_p n_p/c$ is the pump wavenumber, $k_j(\Omega)=(\omega_j+\Omega) n_j/c$ are the wavenumbers associated to signal/idler photons, $n_{p,s,i}$ are refractive index, $L$ is the length of the nonlinear crystal, $G$ is the parametric gain of the nonlinear process.

Idler photons traverse one arm of the interferometer with length $L_i$, probe a sample with reflectivity $r_{i}$ and are reflected back towards the nonlinear crystal where the second parametric interaction takes place. The quantum description of the interaction with the sample can be described as  
\begin{equation}
a_{i_1}(\Omega) \Longrightarrow r_{i}(\Omega) a_{i_1}(\Omega)+f_{i}(\Omega)
\label{loss1}
\end{equation}
where the quantum operator $f_i$ fulfills the commutation relationship $\left[ f_i (\Omega_1), f_i^{\dagger}(\Omega_2) \right] = \left[ 1-|r_i (\Omega_1)|^2 \right] \delta \left(\Omega_1-\Omega_2 \right)$. Signal photons traverse another arm of the interferometer (length $L_s$) before being reflected back towards the nonlinear crystal. Any losses in the signal path can be taken into account considering
\begin{equation}
a_{s_1}(\Omega) \Longrightarrow r_{s} a_{s_1}(\Omega)+f_{s}(\Omega) \label{loss2}
\end{equation}
where the quantum operator $f_{s}$ fulfills the commutation relationship $\left[ f_{s}(\Omega_1), f_{s}^{\dagger}(\Omega_2) \right]= \left[ 1-|r_{s}|^2 \right] \delta \left( \Omega_1-\Omega_2 \right)$. We assume that signal photons experience a reflectivity $r_s$ that shows no frequency dependence in the frequency range of interest here.

After traversing the two arms of the interferometers, signal and idler photons are injected back in the nonlinear crystal where the second parametric amplification takes place, mediating the generation of signal $s_2$ and idler $i_2$ photons. The output signal $a_{s_2}$ and idler $a_{i_2}$ quantum operators can be written as
\begin{eqnarray}
& & a_{s_2}(\Omega)=U_{s_2} a_{s_1} (\Omega)+V_{s_2} a_{i_1}^{\dagger}(-\Omega) \nonumber \\
& & a_{i_2}=U_{i_2} a_{i_1} (\Omega)+V_{i_2} a_{i_1}^{\dagger} (-\Omega) \label{bogoliuvov2}
\end{eqnarray}
We can also take into account detection losses by making the transformation 
$a_{s_2}(\Omega) \Longrightarrow \sqrt{T_s}\, a_{s_2} (\Omega)+g_{s} (\Omega)$
where the quantum operator $g_{s}$ fulfills the commutation relationship $\left[ g_{s}(\Omega_1), g_{s}^{\dagger}(\Omega_2) \right]= \left[ 1-T_{s} \right] \delta \left(\Omega_1-\Omega_2 \right)$. 

Making use of Eqs.~\eqref{bogoliuvov1}, \eqref{loss1}, \eqref{loss2}, and \eqref{bogoliuvov2}, we obtain that the operator $a_{s_2}(\Omega)$ reads

\begin{align}
a_{s_2}(\Omega)
&= A_s b_s(\Omega) + B_s f_s(\Omega) \nonumber\\
&\quad + C_s b_i^{\dagger}(-\Omega)
+ D_s f_i^{\dagger}(-\Omega)
+ g_s(\Omega).
\label{signal_s2}
\end{align}

where

\begin{align}
A_s
&= \sqrt{T_s}\Big\{
r_s(\Omega) U_{s_1}(\Omega) U_{s_2}(\Omega)
\exp\!\left[i\varphi_s(\Omega)\right] \nonumber\\
&\quad
+ r_i^*(-\Omega) V_{s_2}(\Omega) V_{i_1}^*(-\Omega)
\exp\!\left[-i\varphi_i(-\Omega)\right]
\Big\},
\nonumber\\[2pt]
B_s
&= \sqrt{T_s}\,U_{s_2}(\Omega),
\label{parameters}
\\[2pt]
C_s
&= \sqrt{T_s}\Big\{
r_s(\Omega) U_{s_2}(\Omega) V_{s_1}(\Omega)
\exp\!\left[i\varphi_s(\Omega)\right] \nonumber\\
&\quad
+ r_i^*(-\Omega) V_{s_2}(\Omega) U_{i_1}^*(-\Omega)
\exp\!\left[-i\varphi_i(-\Omega)\right]
\Big\},
\nonumber\\[2pt]
D_s
&= \sqrt{T_s}\,V_{s_2}(\Omega).
\nonumber
\end{align}

$\varphi_s(\Omega)=(\omega_s+\Omega) L_s/c$, $\varphi_i(\Omega)=(\omega_i+\Omega) L_i/c$ and $L_{s,i}$ are the lengths of the arms of the interferometer traversed by signal and idler photons, respectively.

We use a laser at the idler wavelength to seed the first parametric amplification process, while the input quantum state of signal photons is the vacuum. If we model the quantum state of the laser light as a multimode coherent state with amplitude $\alpha(\delta \omega+\Omega)$, the input quantum state is $|\Psi_{in} \rangle=|0 \rangle_s\,|\alpha(\Omega) \rangle_i$. In our experiments we observe a small central frequency mismatch $\delta \omega=\omega_i-\omega_L$, where $\omega_L$ is the central frequency of the laser light ($\omega_L$).

The frequency spectrum $S(\Omega)=\langle N_{s_2}(\Omega) \rangle$ of output signal photons $s_2$ is

\begin{equation}
\begin{aligned}
S(\Omega)
&= \left\langle
a_{s_2}^{\dagger}(\Omega)\,
a_{s_2}(\Omega)
\right\rangle \\
&= S_0(\Omega)
\left[1-\left|r_i(-\Omega)\right|^2\right] \\
&\quad + S_0(\Omega)
\left[1+\left|\alpha(\delta\omega+\Omega)\right|^2\right]
\Big[|r_s|^2+\left|r_i(-\Omega)\right|^2 \\
& \quad + r_s r_i(-\Omega)e^{i\Phi}
+r_s r_i^*(-\Omega)e^{-i\Phi} \Big].
\end{aligned}
\label{spectrum1}
\end{equation}
where

\begin{equation}
S_0(\Omega)= T_s\, G^2\,  \text{sinc}^2 \left( \frac{\Delta k(\Omega) L }{2} \right)\, 
\label{spectrum2}
\end{equation}
and
\begin{equation}
\Phi=\frac{\omega_s}{c} \left( L+L_s \right)+\frac{\omega_i}{c} \left( L_i+L \right) + \Delta \varphi_p+\frac{\Omega}{c} \left[ \Delta L + N_g L \right] 
\label{phase}
\end{equation}
$\varphi_p=\varphi_{p_1}-\varphi_{p_2}$ is the phase difference between the pump beams that pump the two parametric amplification processes, $\Delta L=L_s-L_i$ is the path imbalance between the signal and idler arms of the SU(1,1) interferometer, $N_g=N_s-N_i$ is the group index mismatch between signal and idler photons inside the nonlinear crystal. 

In our experiments, we consider objects composed of several layers.  We model these objects as $M$ layers ($M+1$ boundaries between layers), where each layer has thickness $d_i$, refractive index $n_i$ and group index $n_{g,i}$. There are $M+1$ reflections, and we designate the reflection coefficient from the first layer as $R_0=(n_1-1)/(n_1+1)$. The reflectivity that we need to consider in Eq.~\eqref{spectrum1} can be written as

\begin{equation}
r_i(\Omega)=R_0+\sum_{k=1}^M R_k \exp \left[ 2i \sum_{j=1}^k \left( \frac{\omega_i}{c} n_j+ \frac{\Omega}{c} n_{g,j} \right)\,d_j \right]
\end{equation}

where $R_k=\left(n_{k+1}-n_k \right)/ \left( n_{k+1}+n_k)\right)$

\section*{Supplement 3: Conditions on the path length difference \texorpdfstring{$\Delta L$} to perform high sensitivity Fourier-domain OCT}

For the sake of simplicity, let us assume that there is no object in the idler path, $|r_i(-\Omega)|=1$, and that $|r_s|^2=1$. From Eq.~\eqref{spectrum1}, we can derive that the spectrum of the output signal photons $s_2$ is  
\begin{equation}
\begin{aligned}
S(\Omega)
&= 2S_0(\Omega)
\Bigg\{
1+\cos\Bigg[
\frac{\omega_s}{c}(L_s+L)
+\frac{\omega_i}{c}(L_i+L) \\
&\qquad
+\Delta\varphi_p
+\frac{\Omega}{c}[\Delta L+N_gL]
\Bigg]
\Bigg\}.
\end{aligned}
\label{spectrum22}
\end{equation}
There are fringes in the spectrum with a period $\Delta \Omega=2 \pi c/\left( \Delta L + N_g L \right)$. 

On the one hand, in order to perform FD-OCT, we need to be able to measure changes of the fringes, since they will be modified due to the presence of an object in the idler path. Therefore, we need that the bandwidth $\Delta \lambda_s$ of $S_0$ is larger than the period of the fringes, so $\Delta L +N_g L \gg \lambda_s^2/\Delta \lambda_s$.

On the other hand, to resolve with good accuracy the shape of the fringes, we need that the spectral resolution $\delta \lambda_s$ of the spectrometer is smaller than the period of the fringes, therefore $\Delta L + N_g L \ll \lambda_s^2/\delta \lambda_s$. Combining both conditions, we obtain Eq.~(2) in the main text.

\section*{Supplement 4: Sensitivity decay correction}

\begin{figure*}[t!]
    \centering
    \includegraphics[width=1\linewidth]{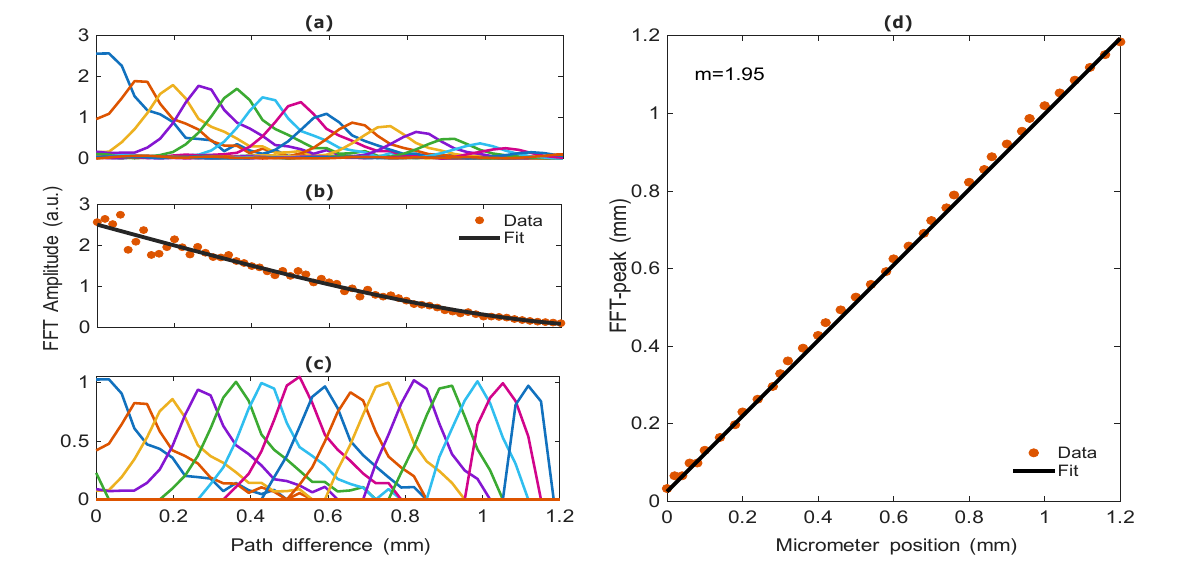}
    \caption{(a) FFT of the measured spectrum. (b) Position of the maximum of each FFT and a fit using the equation $Y_{fit}(x)=A~\text{exp}(-Bx^2)~\text{sinc}^2(Cx)$. (c) Peaks after applying the correction, and (d) relationship between the recovered position from the FFT and the position of the signal mirror. }
    \label{fig:decay}
\end{figure*}

In Fourier-domain OCT, the detected signal amplitude decreases with increasing optical path difference due to the finite spectral resolution of the detection system. This effect, commonly referred to as sensitivity decay or roll-off, arises from the wavelength-dependent resolution of the spectrometer and leads to a systematic attenuation of deeper features in the reconstructed depth profile. To enable accurate comparison of signal amplitudes across different depths, a sensitivity decay correction was applied to selected measurements.

The correction function was experimentally determined by placing a mirror in the sample arm and recording the idler spectrum for different positions of the mirror along the axial direction. For each position, the recorded spectrum was Fourier transformed, and the amplitude of the resulting FFT peak was extracted. Figure~\ref{fig:decay}(a) shows a representative FFT of the measured spectrum, where the single peak corresponds to the optical path difference introduced by the mirror displacement.

The peak amplitudes obtained from multiple mirror positions were then plotted as a function of depth, as shown in Fig.~\ref{fig:decay}(b). The observed decay was fitted using an empirical roll-off function that captures the response of the spectrometer. The inverse of this fitted function was subsequently used to correct the FFT amplitudes obtained in OCT measurements.

Figure~\ref{fig:decay}(c) shows the FFT peaks after applying the sensitivity decay correction, demonstrating uniform peak amplitudes across the accessible depth range. Finally, Fig.~\ref{fig:decay}(d) compares the recovered axial positions obtained from the corrected FFTs with the actual mirror positions, confirming a linear relationship and validating the accuracy of the correction procedure.

\section*{Supplement 5: Sample QEL fabrication}
The structures were fabricated by optical lithography using a Heidelberg Maskless Aligner equipped with a $532~\text{nm}$ laser. Prior to photoresist coating, the samples were cleaned, treated with oxygen plasma, and dehydrated to improve photoresist adhesion. AZ ECI3007 positive photoresist was then spin-coated and patterned by laser exposure. The samples were developed for 1 min, and the cavity profiles were inspected to verify their successful definition. Silver was subsequently deposited by magnetron sputtering under a constant argon flow. Finally, the lift-off process was performed by immersing the samples in acetone and sonicating them for 3~min to remove the remaining photoresist and any residual material.

\section*{Supplement 6: Comparison of the sensitivities in the estimation of reflectivity between unseeded OCT in the high parametric gain regime and seeded OCT in the low parametric gain regime}
For the sake of simplicity, let us consider the single mode approximation, equivalent to considering $\Omega=0$ in Eqs.~\eqref{signal_s2} and \eqref{parameters}. 
\subsection*{Derivative with respect to the reflectivity
\texorpdfstring{$R_i$}{Ri}
of the mean number of signal photons
\texorpdfstring{$\langle N_{s_2}\rangle$}{mean signal photon number}}
We can write that the mean number of signal photons $s_2$  is
\begin{equation}
\langle N_{s_2} \rangle=a+b \left[ 1+|\alpha|^2 \right]+d \left[ 1+|\alpha|^2 \right]\, \cos \Phi
\end{equation}
where
\begin{eqnarray}
& & a =T_s\, \left[1-R_i\right]\,|V|^2 \nonumber \\
& & b= T_s\,\left[ R_s+R_i \right]\,|U|^2 |V|^2 \nonumber \\
& &  d =2 T_s\sqrt{R_s R_i}\,|U|^2|V|^2. \label{parameters2}
\end{eqnarray}
The derivative of $\langle N_{s_2} \rangle$ with respect the reflectivity $R_{i_1}$ that we want to estimate, reads as
\begin{equation}
\frac{d\langle N_{s_2} \rangle}{d R_{i_1}}=p+q \cos \Phi
\label{mean10}
\end{equation}
where
\begin{eqnarray}
& & p =-T_s\,|V|^2 + T_s\,|U|^2 |V|^2\,\left[ 1+|\alpha|^2 \right]  \nonumber \\
& &  q =T_s \sqrt{\frac{R_s}{R_i}}\,|U|^2|V|^2\, \left[ 1+|\alpha|^2 \right] 
\end{eqnarray}

\subsection*{Calculation of the variance of the quantum operator \texorpdfstring{$N_{s_2}$}{Ns2}}
The variance is
\begin{equation}
\text{Var} \left( N_{s_2} \right)=\langle N_{s_2}^2 \rangle-\langle N_{s_2} \rangle^2
\end{equation}
The value of $ \langle N_{s_2}^2 \rangle$ is

\begin{equation}
\begin{aligned}
\langle N_{s_2}^2\rangle&=\left\langle a_{s_2}^{\dagger}a_{s_2} a_{s_2}^{\dagger}a_{s_2} \right\rangle\\
&=\Big\langle\left(A_s^*b_s^{\dagger}+B_s^*f_s^{\dagger}+C_s^*b_i+D_s^*f_i+g_s^{\dagger}\right)\\
&\quad\times\left(A_sb_s+B_sf_s+C_sb_i^{\dagger}+D_sf_i^{\dagger}+g_s\right)\\
&\quad\times
\left(
A_s^*b_s^{\dagger}
+B_s^*f_s^{\dagger}
+C_s^*b_i
+D_s^*f_i
+g_s^{\dagger}
\right)
\\
&\quad\times
\left(
A_sb_s
+B_sf_s
+C_sb_i^{\dagger}
+D_sf_i^{\dagger}
+g_s
\right)
\Big\rangle\\
&=
|A_s|^2|C_s|^2
(1+|\alpha|^2)
+
|B_s|^2|C_s|^2
(1+|\alpha|^2)
(1-R_s)\\
&\quad+
|C_s|^4
(1+3|\alpha|^2+|\alpha|^4)
+
|C_s|^2
(1+|\alpha|^2)
(1-T_s)\\
&\quad+
|A_s|^2|D_s|^2
(1-R_i)
+
|B_s|^2|D_s|^2
(1-R_{s_1})
(1-R_i)
\\
&\quad+
|C_s|^2|D_s|^2|\alpha|^2
(1-R_i)
+
|D_s|^4
(1-R_i)^2
\\
&\quad+
|D_s|^2
(1-R_i)
(1-T_s)\\
&\quad +2|C_s|^2|D_s|^2
(1+|\alpha|^2)
(1-R_i).
\label{N2}
\end{aligned}
\end{equation}

By means of a straightforward but somehow cumbersome calculation, we obtain that 

\begin{align}
\text{Var} \left( N_{s_2} \right) & = |C_s|^2 \left[1+|\alpha|^2 \right] \{ |A_s|^2+|B_s|^2 \left[1-R_s \right] +\left[1-T_s \right] \} \nonumber\\
& +|D_s|^2 \left[1-R_i \right] \{ |A_s|^2+|B_s|^2 \left[1-R_s \right] +\left[1-T_s \right] \} \nonumber\\
& \quad |\alpha|^2 |C_s|^4 +|C_s|^2|D_s|^2 \left[1-R_i \right] |\alpha|^2 
 \label{variance1}
\end{align}

If we make use of the commutation relationship 
\begin{align}
[a_{s_2},a_{s_2}^{\dagger}]&=|A_s|^2+|B_s|^2 \left( 1-R_s \right) \nonumber\\
& + \left( 1-T_s \right)-|C_s|^2-|D_s|^2 \left( 1-R_i \right)=1,
\end{align}

so we can write
\begin{equation}
|A_s|^2+|B_s|^2( 1-R_s)+( 1-T_s)=1+|C_s|^2+|D_s|^2 ( 1-R_i),
\end{equation}
the variance given by Eq.~\eqref{variance1} can be simplified to
\begin{align}
\text{Var} \left( N_{s_2} \right) &= \Big\{ |C_s|^2+|D_s|^2| \left[1-R_i \right] \Big\} \nonumber \\
& \quad \times \Big\{ 1+|C_s|^2+|D_s|^2| \left[1-R_i \right] \Big\}   \nonumber \\
&  \quad + |\alpha|^2 |C_s|^2 \Big\{  1+2|C_s|^2+2|D_s|^2 \left[1-R_i \right] \Big\} 
\end{align}

Making use of Eqs.~\eqref{parameters} and \eqref{parameters2}, we can write $|C_s|^2=b+d\cos \Phi$ and $|D_s|^2 \left[1-R_i \right]=a$, so 
\begin{equation}
\text{Var} \left( N_{s_2} \right)=A+B \cos \Phi+D \cos^2 \Phi
\label{variance10}
\end{equation}
where
\begin{eqnarray}
     & & A=(a+b)(1+a+b)+|\alpha|^2 b(1+2a+2b), \nonumber \\
     & & B=d(1+2a+2b)+|\alpha|^2 d(1+2a+4b), \\
     & & D=d^2(1+2|\alpha|^2), \nonumber 
\end{eqnarray}
\subsection*{The precision in the estimation of \texorpdfstring{$R_i$}{Ri}}
The precision $\sigma^2$ of the estimation of $R_i$ by measuring $\langle N_{s_2} \rangle$ is given by the propagation of errors equation as
\begin{equation}
\sigma^2=  \frac{\text{Var} \left( N_{s_2} \right)}{\Big[ \partial \langle N_{s_2}  \rangle/\partial R_i \Big]^2 }
\end{equation}
Making use of Eqs.~\eqref{mean10} and \eqref{variance10}, we obtain that
\begin{equation}
\sigma^2=  \frac{A+B \cos \Phi+D \cos^2 \Phi}{p^2+2pq \cos \Phi+q^2 \cos^2 \Phi}
\end{equation}
Minimization of the variance shows that the optimal operating phase is $\Phi=\pi$. Therefore, the minimum variance is
\begin{equation}
\sigma^2=  \frac{A-B+D }{\left(p-q \right)^2}
\label{minimum variance}
\end{equation}
Let us first consider the case of a constant value of the gain (low-parametric gain regime), and increasingly larger values of the seed ($|\alpha|^2 \gg 1)$. In this case, the minimum sensitivity tends to
\begin{align}
\sigma_{min}^2
&=
\frac{1}{|\alpha|^2}\,
\frac{R_s+R_i-2\sqrt{R_sR_i}}
{T_s|U|^2|V|^2\,
\left(1-\sqrt{R_s/R_i}\right)^2}
\nonumber\\
&\qquad\times
\left\{
1+2T_s(1-R_i)|V|^2
\right.
\nonumber\\
&\qquad\left.
+2T_s\,
\left[R_s+R_i-2\sqrt{R_sR_i}\right]
|U|^2|V|^2
\right\}.
\label{sigmamin}
\end{align}

In the second scenario, we consider increasingly larger values of the gain gain ($ G \gg 1$) and no seed ($|\alpha|^2=0$). In this case, the minimum variance given by Eq.~\eqref{minimum variance} stays constant for increasingly larger values of the gain with a value of
\begin{equation}
\sigma_{min}^2=\frac{\left( R_s+R_i\right)^2-4 \left( R_s+R_i\right)\, \sqrt{R_s\,R_i}+4 R_s\,R_i}{\left[ 1-\sqrt{R_s
/R_i}\right]^2}
\end{equation}

\bibliography{references}

\end{document}